\documentstyle[epsfig,floats,twocolumn,prl,aps]{revtex}
\input epsf
\input rotate

\begin{document}
\title{ Linear Stability of Scroll Waves}
\author{Herv\'e Henry and Vincent Hakim}
\address{ Laboratoire de Physique Statistique,
Ecole Normale Sup\'erieure,\\ 24 rue Lhomond, 75231 Paris Cedex 05, France}
\date{\today}
\maketitle
\begin{abstract}
A full linear stability of a straight 
scroll wave in an excitable
medium is presented. The five eigenmode branches  which correspond to
deformation in the third 
dimension of the five main modes of two-dimensional (2D)
spiral dynamics are found to play a 
dominant role. Modulations
in the third dimension  have  stabilizing or destabilizing
effects on the different modes depending on the parameter regimes. For untwisted
scroll waves, our numerical results confirm the relation between
the long-wavelength behavior of the translation branches and
2D spiral drift in an external field but show no similar direct
relation for the
meander branches. The influence of twist on the different
branches is investigated. 
 In particular,
the sproing instability is seen 
to arise from the twist induced deformation of the 
translation
branches above a threshold twist.
\end{abstract}

Scroll waves are three-dimensional (3D) extensions of
the familiar spirals \cite{Win0} of excitable media. They have been
directly observed in three-dimensional chemical reactions in gels
\cite{pert} and are suspected to play a role in the slug phase of slime 
molds life cycle \cite{dicty} and most importantly in
ventricular fibrillation
\cite{win3dh,fenkar}. Numerous simulations of scroll waves
have been performed, different
instabilities have been noted\cite{win3d,bik3d,armit} and
a somewhat similar phenomenology has been found for 3D complex Ginzburg-Landau
vortices \cite{cgle}. However, even in the simplest case of an 
homogeneous and isotropic excitable medium, the mechanisms of the different
instabilities, their systematics and the influence of scroll wave twist still 
appear poorly
understood. We attempt here to bring some clarification by performing a full
linear stability analysis of straight scroll waves in various parameter 
regimes. In contrast to earlier
stability studies \cite{armit,garfin}, this gives access not only to the most
unstable eigenmode but also to the other, stable or unstable, modes. 
In particular, this enables us to follow the five branches of modes which
emerges from 
the two translation modes, the rotation zero mode and
the two Hopf bifurcation meander modes of 2D spiral motion. As for spirals
\cite{barkl}, 
these are  found to play the dominant role in scroll wave dynamics.
 
The excitable medium dynamics is described in a simplified but usual way by
the two-variables reaction-diffusion system
\begin{eqnarray}
\partial_t u&=& \nabla^2u+f(u,v)/\varepsilon\label{eq1}
\\
\partial_t v&=&\delta \nabla^2 v+ g(u,v)
\label{eq2}
\end{eqnarray}
We restrict ourselves to the singly diffusive case with $\delta=0$ and 
choose
for definiteness Barkley's reaction terms \cite{barkn}
with $f(u,v)=u(1-u)[u-(v+b)/a],\ g(u,v)=u-v$ which permits fast direct
simulations and comparison with some previous results. Previous experience
with 2D spirals lead us to expect that similar results would be 
obtained with a different choice of excitable kinetics.
The results should also be, at least
qualitatively, comparable to experiments using
Belousov-Zhabotinsky reaction \cite{fless}.

Our numerical method is analogous to the one used in
 a previous
linear stability analysis of spiral waves \cite{barkl}. 
We 
introduce a rotating coordinate system twisted along the $z$-axis where
the steady scroll solution is time independent. That is, we seek $u$ and
$v$ as  functions of $r, \phi=\theta-\omega t -\tau_w z, t$ and $z$ where
$(r,\theta,z)$ are
cylindrical coordinates and $\tau_w$ is the imposed twist. 
In these coordinates, Eq.~(\ref{eq1},\ref{eq2}) read
\begin{eqnarray}
(\partial_t +2\tau_w\partial^2_{\phi z} -\partial^2_{zz})u&=&(\omega
\partial_\phi+
\tau^2_w\partial^2_{\phi\phi}+\nabla^2_{2D})u+f(u,v)/\varepsilon
\nonumber\\
\partial_t v&=&\omega\partial_\phi v+g(u,v)
\label{tdpolar}
\end{eqnarray}
The Laplacian is two dimensional on the r.h.s of Eq.~(\ref{tdpolar})
 [in $(r,\phi)$]
in contrast to the three dimensional one in Eq.~(\ref{eq1}).  The 
twist $\tau_w$ can be chosen at will and is not constrained by
the finite size of the simulation box in the $z$-direction as in usual
direct simulations.
Determining a steady scroll rotating at the frequency $\omega=\omega_1$
consists in finding $(u_0(r,\phi),\ v_0(r,\phi),\ \omega_1)$, that
nullifies the r.h.s of Eq.~(\ref{tdpolar}).
To solve this nonlinear eigenvalue problem, the r.h.s. of 
Eq.~(\ref{tdpolar})
is discretized on a $N_R \times N_\phi$ finite, $2\pi \
\phi-$periodic  box of radius $R$ using finite differences of 
$4^{\mbox{\footnotesize{th}}}$
order in space for the differential operators with
rotationally symmetric boundary conditions imposed at the box edge,
$\partial_r u_0|_{r=R}=0$.
The non-uniqueness of $u_0(r,\phi),\ v_0(r,\phi)$ brought by
the rotational invariance of (\ref{tdpolar}) is
taken care of by arbitrarily setting the value  of
$u_0(N_r,N_\phi)$ equal to 0.5. Therefore the discrete problem is
reduced to finding the zero of a $(2\times N_r \times
N_\phi)$  valued function  of 
the $(2\times N_r \times
N_\phi -1)$ unknown values of $u_0$ and $v_0$ and 
$\omega_1$. This is solved  by Newton's method
starting from an initial guess of the solution provided by a direct
simulation of the evolution equations.
In a box of size $60\times120$, solutions of the discrete
problem are obtained with an 
accuracy of 
$10^{-8}$ (in a $L_2$ norm).

Once a steady scroll wave $(u_0,v_0,\omega_1)$  
is found, the time evolution equations (\ref{tdpolar})
are linearized around $(u_0,v_0)$ in the twisted rotating frame.
Setting $u=u_0+
\exp[\sigma(k_z) t -i k_z z] u_1(r,\phi), 
v=v_0+ \exp[\sigma(k_z) t -i k_z z] 
v_1(r,\phi)$, leads to 
a linear eigenvalue problem for the growth rates
$\sigma(k_z)$  as a function of the wave-vector $k_z$,
\begin{eqnarray}
\sigma(k_z)\, u_1 &=& (-k_z^2+2 i\tau_w k_z\partial_{\phi}) u_1+
(\omega_1\partial_{\phi}+\tau_w^2 \partial_{\phi\phi}^2+ \nabla^2_{2D}
)u_1+\nonumber\\
&+&[\partial_u f(u_0,v_0)u_1+\partial_v f(u_0,v_0)v_1]/\varepsilon
\nonumber\\
\sigma(k_z)\, v_1 &=& \omega_1\partial_{\phi}v_1 +
[\partial_u g(u_0,v_0)u_1+\partial_v g(u_0,v_0)v_1]
\label{eqlin}
\end{eqnarray}
These linear equations are discretized in the same manner as 
Eq.~(\ref{tdpolar}).
We denote by ${\mathcal{L}}_{k_z}$ the discrete version of the linear operator
appearing on the r.h.s. of Eq.~ (\ref{eqlin}).
Being interested in the stability of the scroll wave and in its modes of
destabilization, 
we focus on
determining
the eigenvalues of ${\mathcal{L}}_{k_z}$  of largest real parts. This can
be accurately done by using an
iterative method fully described in \cite{Gold87}. 
Firstly, the contributions of eigenvalues with very negative real parts
are effectively suppressed by 
computing by iterations   
$x_1=(1+dt{\mathcal{L}}_{k_z})^{t_0/dt} x_0$, for an arbitrary vector $x_0$
and
a sufficiently large
integer $t_0/dt$ and a sufficiently small
$dt$ 
(in the following calculations, typical values are $t_0=5$ and
$dt\simeq 10^{-5}$). 
One then builds an orthonormal
set of vectors that spans $\{x_1,\ x_2=(1+dt{\mathcal{L}}_{k_z})^{t_1/dt}x_0,
\ ...,\
x_m=(1+dt{\mathcal{L}}_{k_z})^{(m-1)t_1/dt}x_0\}$ 
and evaluates
the restriction
${\mathcal{L}}_{k_z}^{(m)}$ of ${\mathcal{L}}_{k_z}$ to this subspace. The 
integer 
$t_1/dt$ should be chosen large enough
to make $(1+{\mathcal{L}}_{k_z}dt)^{t_1/dt}$  significantly different
from the identity but small enough to limit computation time (as a typical
value, $t_1=0.5$ is used here).
One finally computes the
$n$ eigenvalues $\{\sigma_j, 1\leq j\leq n \}$ of largest real part 
($Re(\sigma_j)>Re(\sigma_k),j<k$) 
and the corresponding eigenvectors of ${\mathcal{L}}_{k_z}^{(m)}$.
They approximate the corresponding  $n$ eigenvectors of 
${\mathcal{L}}_{k_z}$ 
with an accuracy
of order $\exp [{-(\sigma_n^r-\sigma_m^r)(m-n)t_0}]$. With 
$m=50$, we reach
a precision of more than
$10^{-6}$, in computing eigenvectors of ${\mathcal{L}}_{k_z}$ (in $L_2$ norm
$||[{\mathcal{L}}_{k_z}-\sigma(k_z)](u_1,v_1)||<10^{-6}$).

We have determined steady rotating scroll waves and computed
their stability spectrum for several different
parameters of Eq.(\ref{eq1},\ref{eq2}) where steady spiral waves exist and
are stable.
We now summarize these results beginning with the case of untwisted
scroll waves.

\begin{figure}
\begin{center}
\includegraphics[width=4.cm,height=4.cm]{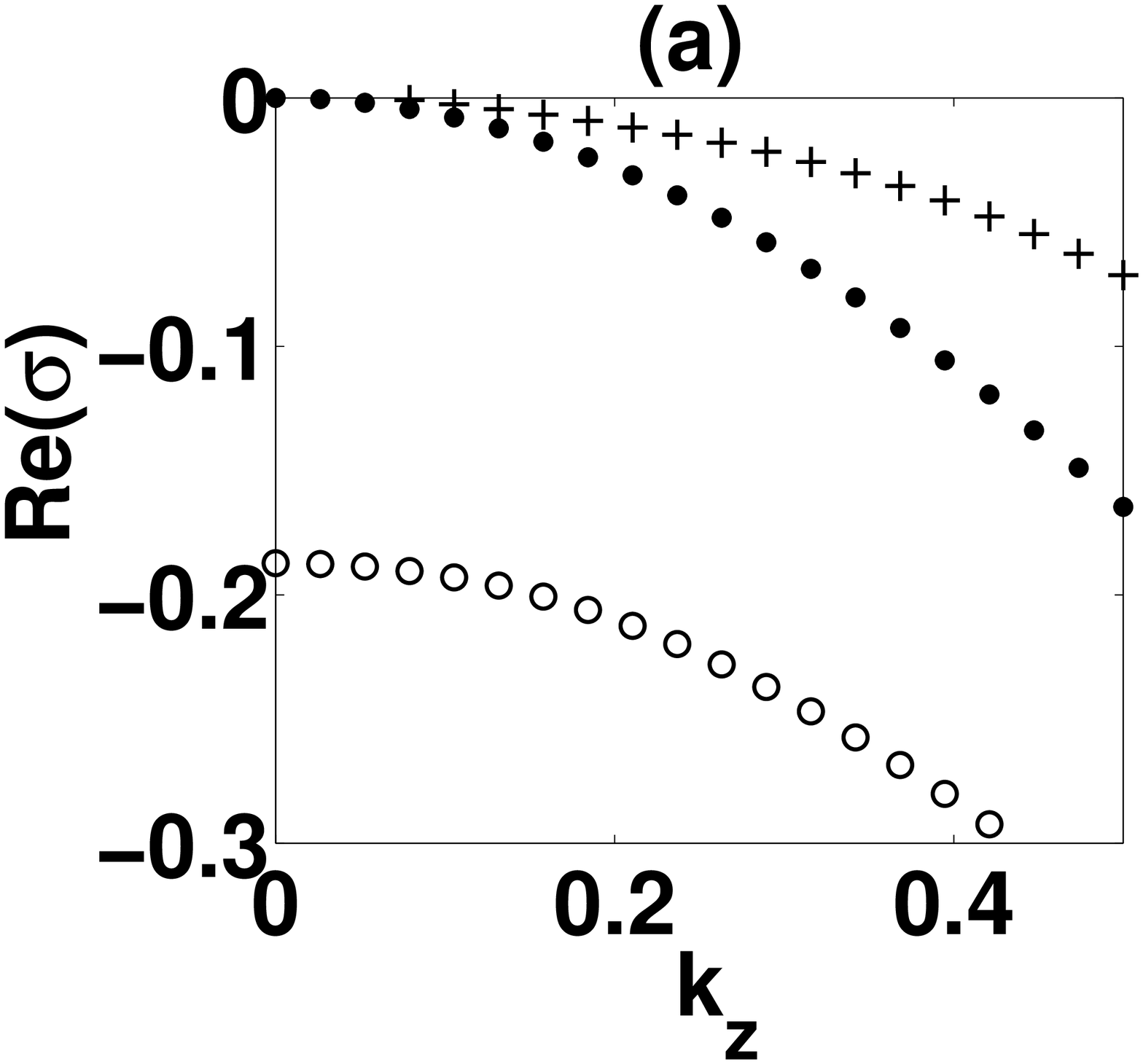}
\includegraphics[width=4.cm, height=4.cm]{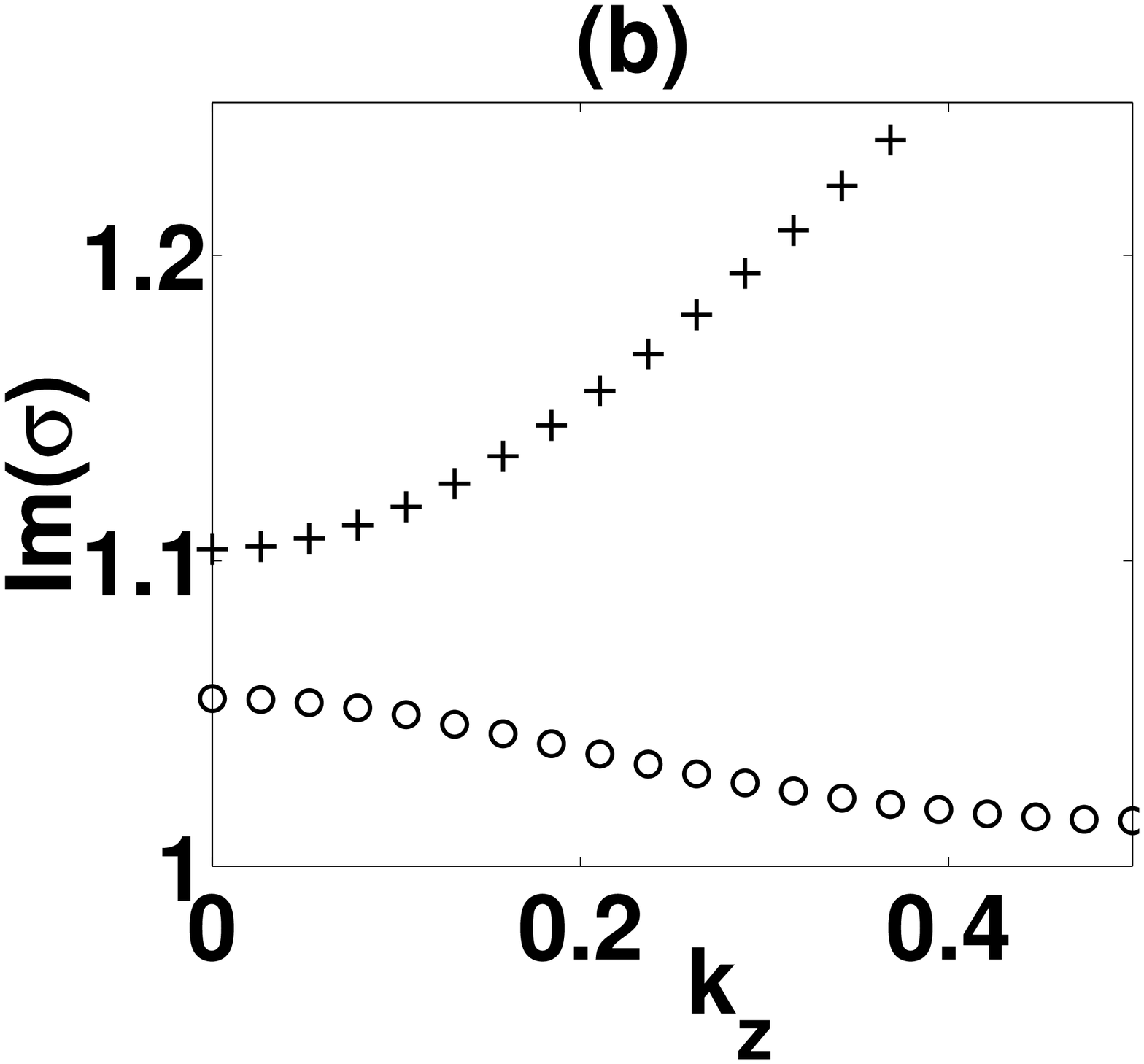}
\caption{
(a) Real and (b) imaginary parts of the dominant
eigenvalues of the spectrum as functions of $k_z$ 
: translation (+), rotation
($\bullet$) and meander($\circ$) branches. 
The kinetics parameters are $a=1.,\ b=.01
\mbox{ and } \varepsilon=.05 $ and qualitatively give 
2D spirals with small core far from the meander line.
No instability is observed.
The spiral rotation frequency is $\omega_1=1.103$ and
the width of the excited zone between two $u=.5$ iso-lines far
from the tip  
corresponds to a wavenumber $k_{\lambda}\simeq 0.90$. A fit for $k_z\ll 1$ 
of the translation branch gives $\sigma_t\simeq 1.1\, i+(-0.38+1.31\, i)\,
k_z^2$ in agreement with the
independently measured drift coefficients 
$\alpha_{\parallel}=0.37,\alpha_{\perp}=1.32$. A similar fit for
the meander branch gives $\sigma_m\simeq -0.18 +1.05 \, i+(-0.49-0.49\, i)\,k_z^2$ }
\label{bothstab}
\end{center}
\end{figure}

\begin{figure}
\begin{center}
\includegraphics[width=4.cm,height=4.cm]{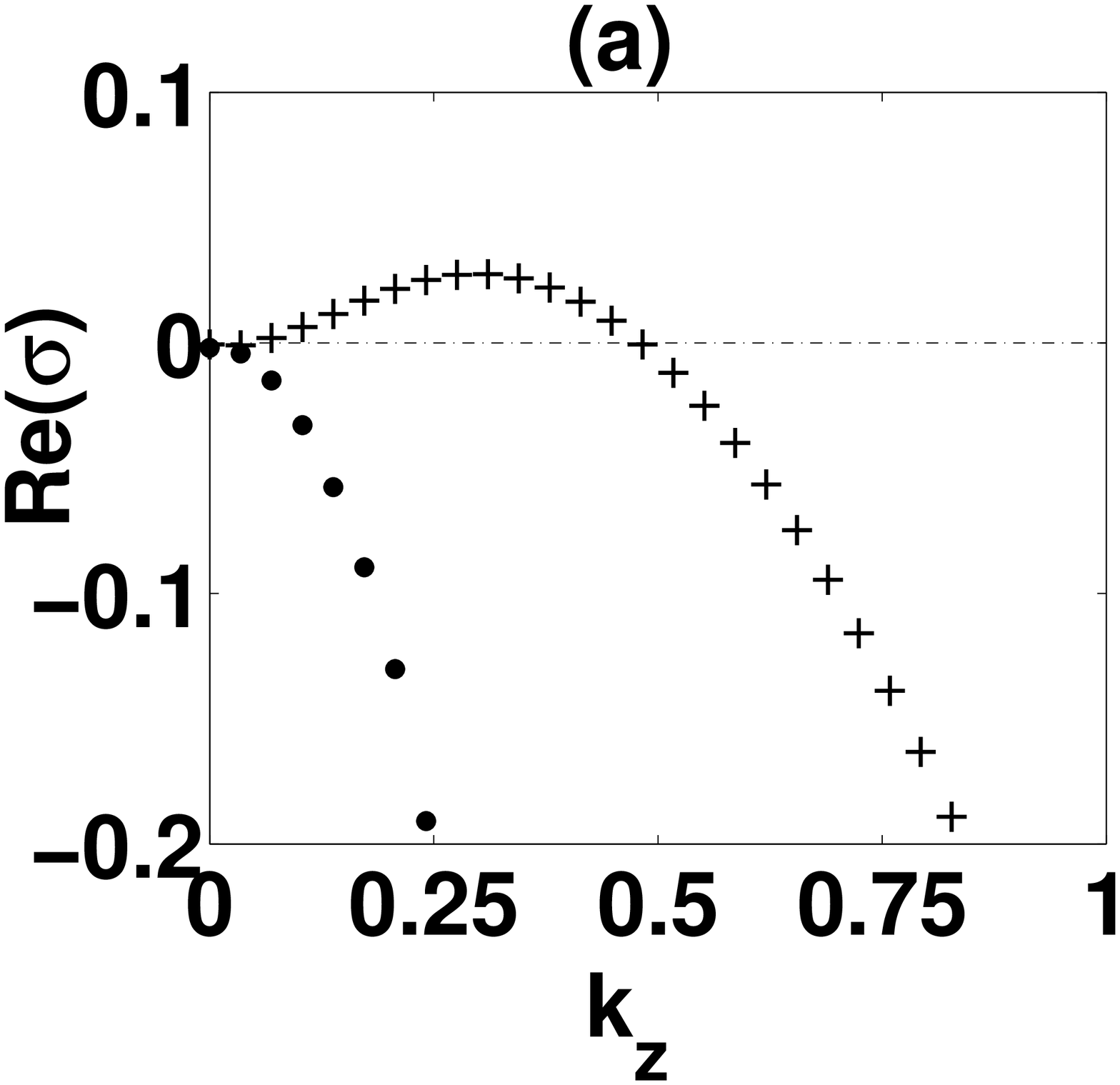}
\includegraphics[width=4.cm, height=4.cm]{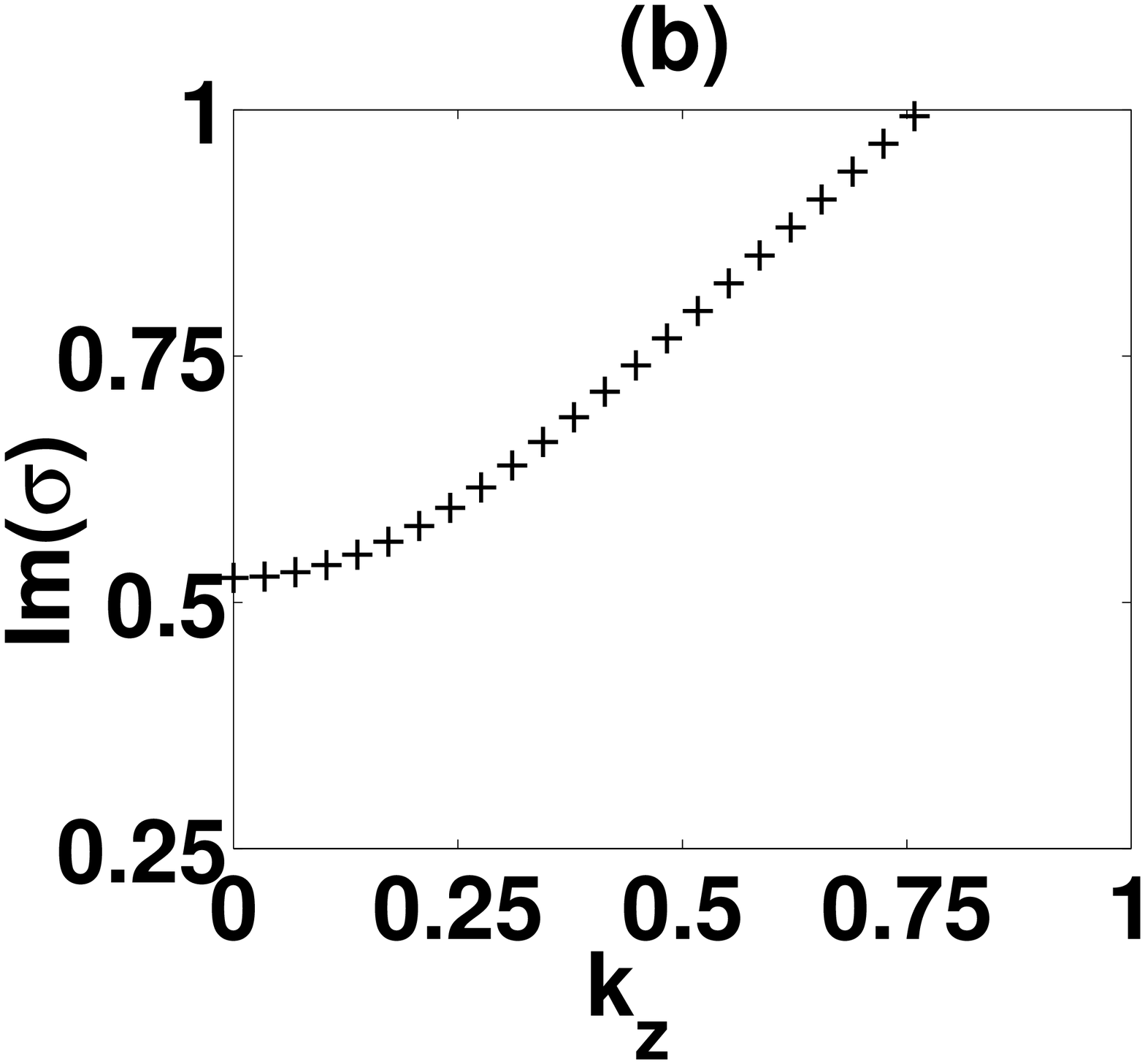}
\caption{
(a) Real and (b) imaginary parts of the dominant
eigenvalues of the spectrum as functions of $k_z$ 
: translation (+) and
rotation ($\bullet$) branches. 
The kinetics parameters are $a=.96,\ b=.001
\mbox{ and } \varepsilon=.1$ and qualitatively give a weakly excitable medium
with large core non-meandering spirals.
The meander
modes are not plotted because their real parts is more negative than the
values shown. The translations branches are unstable for low $k_z$.  
With the notations
of Fig.~\ref{bothstab}, $\sigma_t\simeq .52\, i+(0.88+1.26\, i)\,
k_z^2$ for $k_z\ll 1$ and $\alpha_{\parallel}=-0.87,\alpha_{\perp}=1.25$.
Other values are $\omega_1=0.521,\ k_{\lambda}\simeq 0.95$.} 
\label{transdes}
\end{center}
\end{figure}

\begin{figure}
\begin{center}
\includegraphics[width=4.cm,height=4.cm]{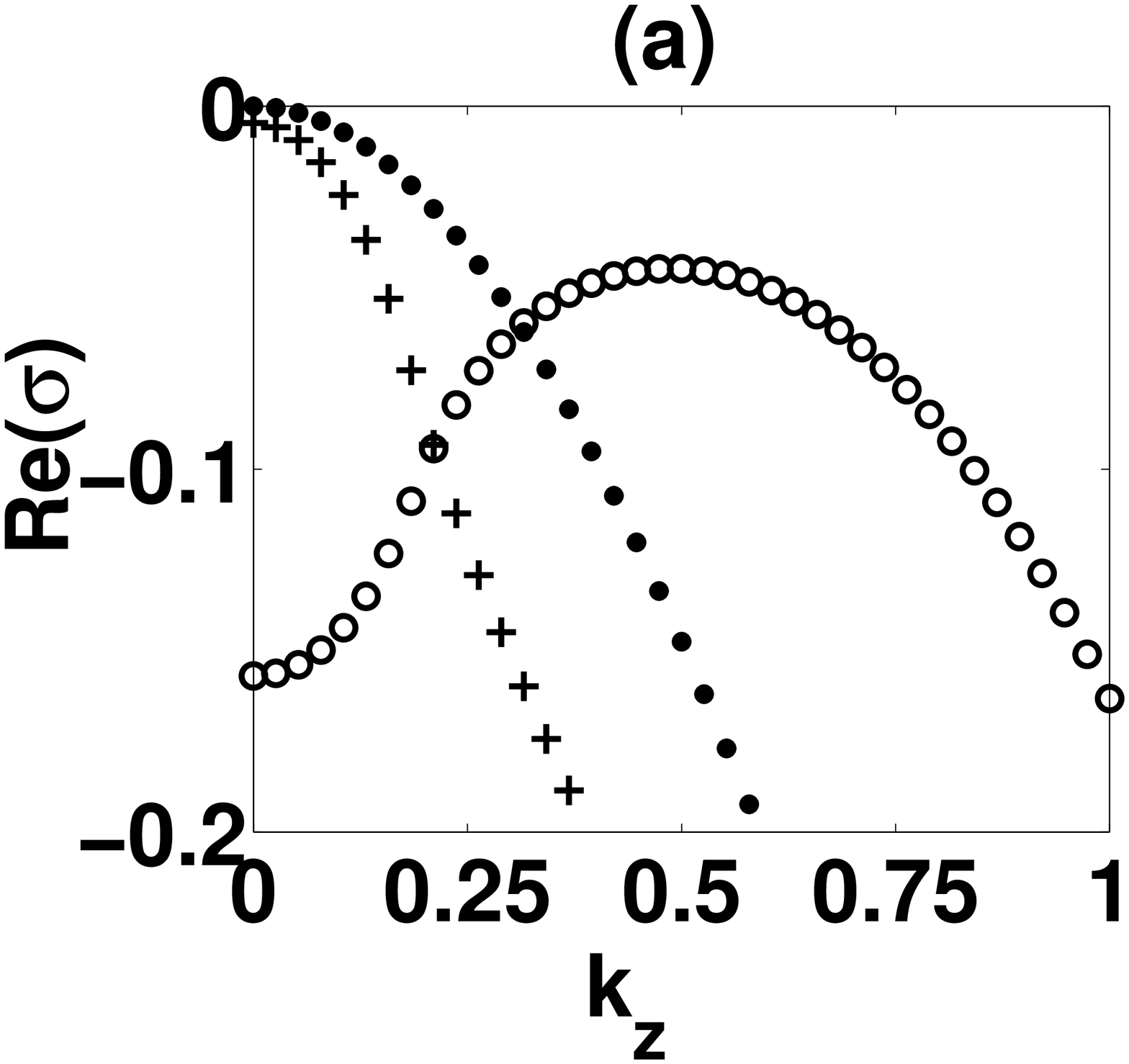}
\includegraphics[width=4.cm, height=4.cm]{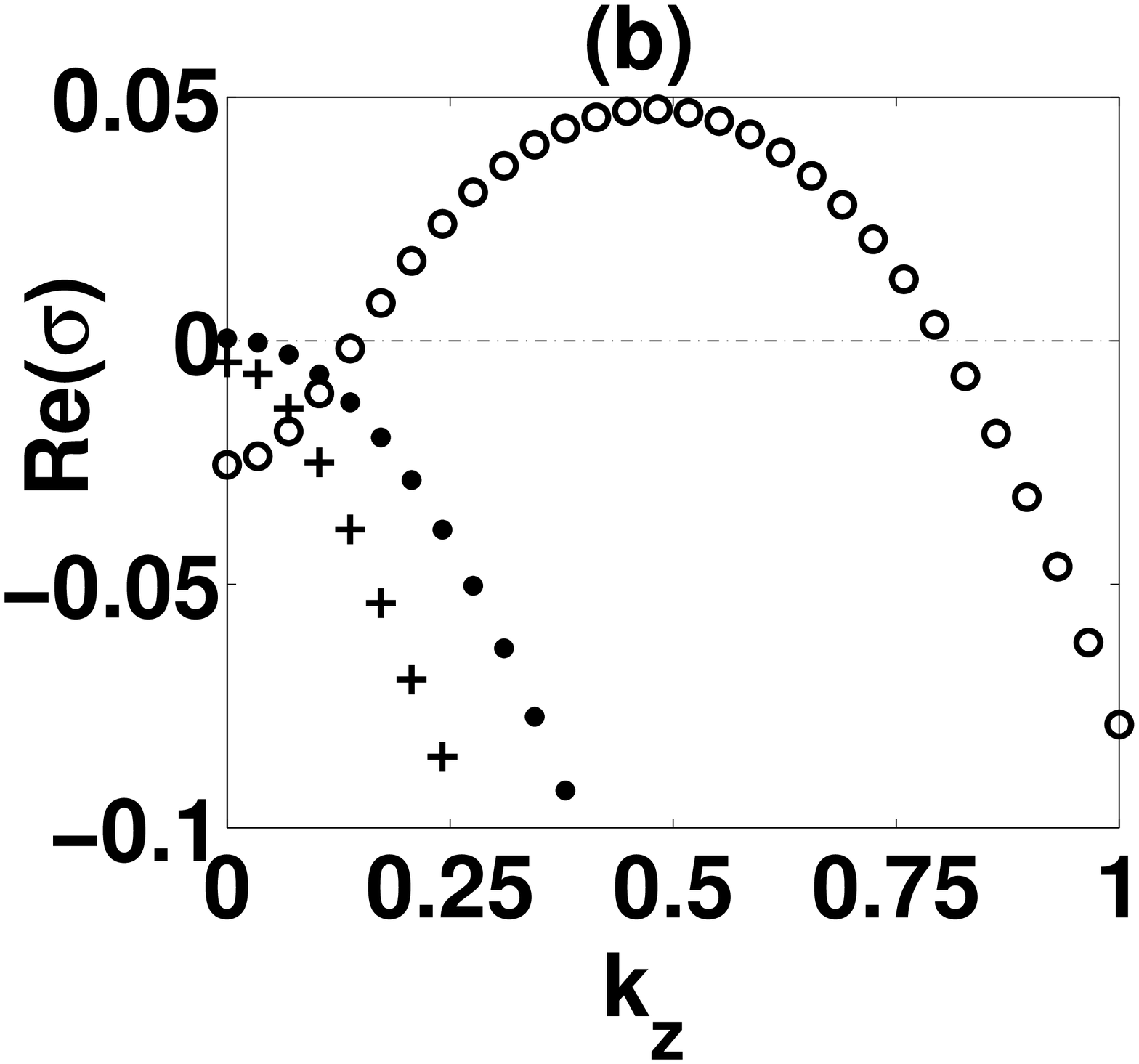}
\caption{Real parts of the dominant
eigenvalues of the spectrum as functions of $k_z$ : 
translation (+), rotation ($\circ$) and
 meander ($\bullet$) branches. The kinetics parameters qualitatively give
small core spirals closer to the 2D meander line in (b) than in (a).
 (a) $a=.7,\ b=.01,$  and 
$\varepsilon=.025$. The value obtained are
$k_{\lambda}\simeq 2.,\omega_1=1.784$ and 
$\alpha_{\parallel}=1.62$, $\alpha_{\perp}=0.83$ to be compared
with
$\sigma_t\simeq 1.78\, i+(-1.63+0.83\, i)\, k_z^2$ and
$\sigma_m\simeq -0.157 +1.94\, i+(1.04+0.21\, i)\, k_z^2$ for $k_z\ll 1$.
(b) $a=.66,\ b=.01, \mbox{ and }    
 \varepsilon=.025$. The value obtained are
$k_{\lambda}\simeq 2.,\ \omega_1=1.756$ and 
$\alpha_{\parallel}=2.44$, $\alpha_{\perp}=0.77$ to be compared
with
$\sigma_t\simeq 1.75 \,i+(-2.41+0.75\, i) k_z^2$ and          
$\sigma_m\simeq -0.071+ 1.89\, i+(1.89+0.33\, i) k_z^2$ for $k_z\ll 1$ . 
In (b) the meander mode 
is stable at $k_z=0$
but the meander branch becomes unstable at finite $k_z$.
}
\label{meandes}
\end{center}
\end{figure}

For 2D-spiral waves, the linear stability operator has three marginally stable
modes
coming from the symmetries of the dynamics : the rotation eigenmode with 
an eigenvalue equal to zero and
the two translation eigenmodes with purely imaginary 
eigenvalues equal to $\pm i\omega_1$, the spiral rotation frequency. For
untwisted scroll waves, these correspond to eigenmodes of the linearized
operator at $k_z=0$. The computations reported in
Figs.~\ref{bothstab},\ref{transdes},\ref{meandes}
show good agreement with these 
expectations. For $k_z=0$, we find an eigenvalue of magnitude
smaller than  $10^{-9}$ which we identify with
the rotation eigenmode  and two eigenvalues equal to  $\pm i
\omega_1$ with an error of $10^{-4}$ which we identify with the translation
eigenmodes \cite{noteprec}.

For 2D-spiral waves, in addition to these three marginal eigenvalues,
two complex
conjugate eigenvalues cross the imaginary axis on the 2D meander instability
line. The corresponding meander eigenmodes play an important role in spiral
dynamics and belong to the untwisted scroll wave spectrum at $k_z=0$.
The results of our computations show that in general,
the eigenvalues with the largest real parts
in the untwisted scroll wave spectrum belong to the five finite-$k_z$ branches
of modes
originating from
the translation, rotation or meander eigenmodes at $k_z=0$. The rotation
mode appear to be stabilized by finite $k_z$ values. On the contrary,
we have observed
three different behaviors 
of the translation and meander branches 
as a function of $k_z$
for different choices of the parameters $a,b,\varepsilon$ characterizing 
the reaction term $f(u,v)/\varepsilon$ in Eq.~(\ref{eq1}):\\
-In the first case, both the translations and the
meander modes are stabilized when $k_z$ becomes non-zero, as shown in 
Fig.~\ref{bothstab}.\\
The second case is shown in Fig.~\ref{transdes}. 
The translation modes now become unstable for low $k_z$ while the meander
modes are stabilized.
This corresponds to the "negative filament tension" instability previously
described in \cite{bik3d}.\\
- A third case opposite to the previous one is also observed as
shown in Fig.~\ref{meandes}. The translations modes
are here stabilized when $k_z$ becomes non-zero. On the contrary,
the real part of the meander modes grows proportionally to $k_z^2$ at
low $k_z$ before decreasing for higher wave numbers. This qualitatively
corresponds to the phenomenon reported in \cite{armit}: the
scroll waves can be unstable with respect to meander at finite $k_z$ in a
parameter regime where 2D spirals are stable, as shown in Fig.~\ref{meandes}b.

The fourth logical possibility consisting of a
destabilization of both the translation and meander modes was not observed
in the present limited search.

We have performed some direct
numerical simulations of Eq.~(\ref{eq1},\ref{eq2}) to observe the nonlinear
development of the different instabilities \cite{barkn}.
When the translation modes are destabilized
at low $k_z$ (Fig.~\ref{transdes}), the filament grows and no restabilization
is observed. On the contrary, a restabilization at finite amplitude 
is observed when the meander modes are destabilized at finite $k_z$. This
3D instability thus inherits the direct character of the 2D meander bifurcation.
These observations corroborates previous findings \cite{bik3d,armit}. However,
a detailed characterization of the nonlinear dynamics 
is beyond the scope of the present letter.

We now proceed and study the effect of a finite twist ($\tau_w\neq 0$)\cite{mb}.
At a general level, we note that the spectra obey $\sigma(-k_z)=\sigma^*(kz)$ 
where the star denotes complex conjugation, since 
${\mathcal{L}}_{-k_z}= {\mathcal{L}}^*_{-k_z}$.
As in the untwisted case, the symmetries of the dynamics
provide three known eigenvalues, one zero eigenvalue at
$k_z=0$ for the rotation eigenmode and two complex
conjugate eigenvalues for the translation eigenmodes:
$i\omega_1$ at $k_z=-\tau_w$ 
corresponding to the eigenvector 
$\exp(i\phi)(\partial_r u_0+i\partial_{\phi} u_0/r,
\partial_r v_0+i\partial_{\phi} v_0/r)$ and $-i\omega_1$ at $k_z=+\tau_w$ for
the complex conjugate eigenvector. 
These exact results provide sensitive checks on the
numerics.

\begin{figure}
\begin{center}
\includegraphics[width=4.cm,height=4.cm]{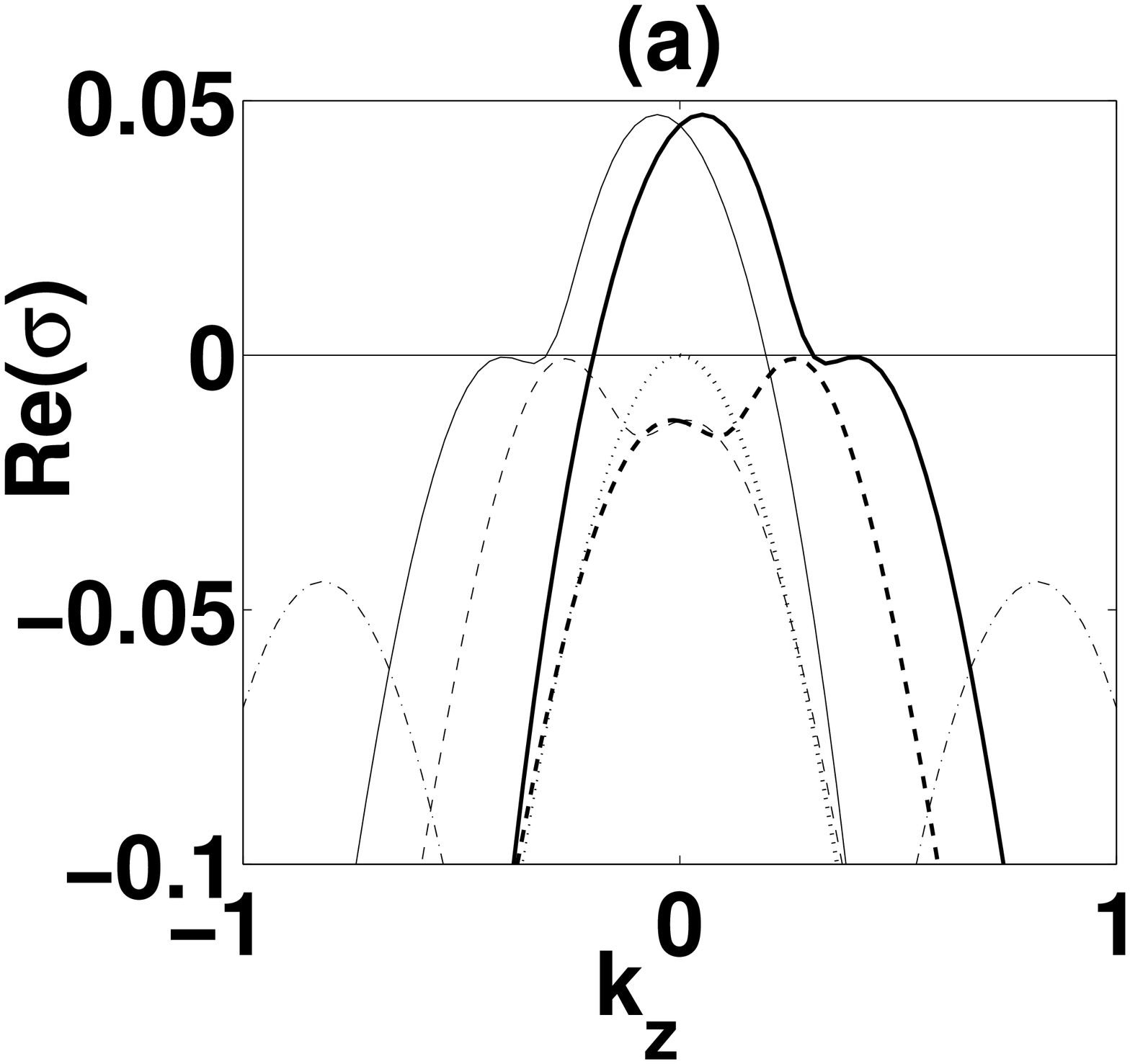}
\includegraphics[width=4.cm, height=4.cm]{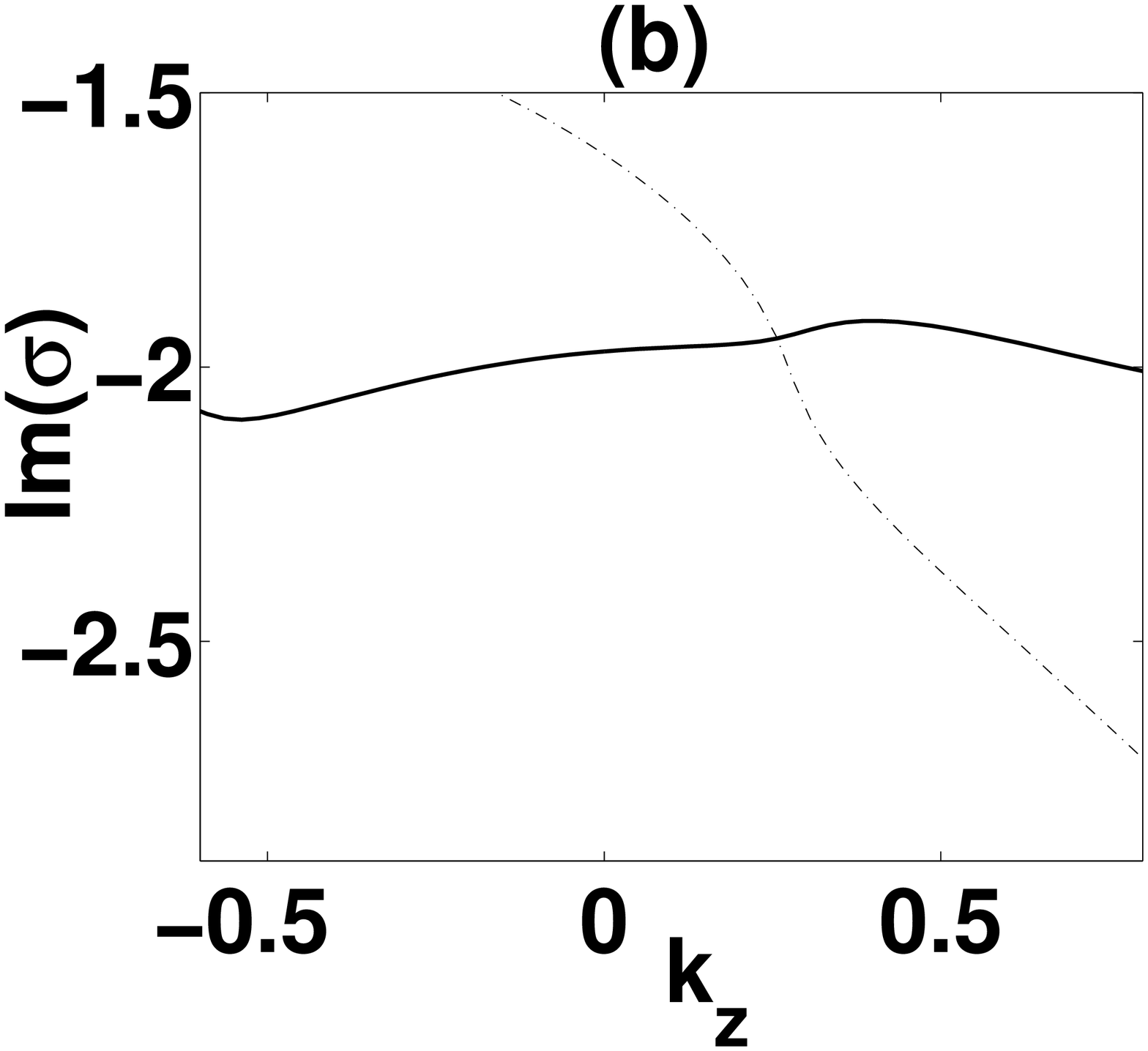}
\caption{\label{twist}(a) Real and (b) imaginary parts of
the dominant eigenmodes for $a =.7$, $ b=0.01$, $\varepsilon=0.025$
(same as Fig.3a)
and a twist $\tau_w=.4$: meander(dashed-dotted), rotation (dotted) and
translation (solid) branches. In (a) one of the translation branch (bold dashed)
is also plotted for a twist $\tau_w=.26$ below the instability
threshold.
}
\end{center}
\end{figure}

We generally find that increasing the twist $\tau_w$ from zero reduces the
stability of the translation branches, destabilize them or amplify the
instability when they are already unstable. The effect of twist
increase appears much less pronounced on the meander modes.
We focus here on the most interesting
parameter regimes such as those of Fig.~\ref{bothstab},\ref{meandes}a
for which the untwisted scroll waves are stable.
A representative case is shown on Fig.~\ref{twist}. The kinetics parameter
are those of Fig.~\ref{meandes}a. When twist is increased from zero, the
zeroes of $Re[\sigma(k_z)]$ which correspond to the translation eigenmodes,
move away from $k_z=0$ and stand at $k_z=\pm \tau_w$ as they should.
Most importantly, the translation eigenmodes remain local
maxima of  $Re[\sigma(k_z)]$. 
Beyond a threshold twist ($\tau_w\approx 0.20$) a secondary
maximum appears for $k_z$ close to zero. One
of the translation branch is shown on Fig.~\ref{twist}a for $\tau_w=0.26$
when $Re[\sigma(k_z)]$ is still negative at this secondary maximum (the
other translation branch can be deduced from the one shown by reflection with
respect to the $k_z=0$ axis). When the twist is increased further (past
$\tau_w \approx 0.30$),
$Re[\sigma(k_z)]$ becomes positive at this secondary maximum. The twisted
scroll waves then become unstable for a finite range of wave numbers as shown 
on
Fig.~\ref{twist}a for $\tau_w=0.4$. The rotation and meander branches are also
shown for this value of twist. One sees that the  
real parts of the two 
meander
branches which are superimposed on Fig.~\ref{meandes}a for $\tau_w=0$
have been
split by the twist.
The maxima around $k_z=\pm.48$ for $\tau_w=0$ have 
shifted around $\pm(.48+\tau_w)$ 
but their stability has not been significantly  modified.

Direct numerical simulations show that this twist-induced instability leads
to a restabilized state in which the scroll core itself 
acquires a helical shape. 
 We can thus safely identify the present instability with
the 'sproing' instability of
\cite{win3d} but again a detailed study of the nonlinear state is best deferred
to another publication.

Our results can be partially rationalized on the basis
of previous works but also show the limitations of some
previous theoretical suggestions.
In the untwisted case, it was  noted \cite{hk} that
a weak scroll curvature 
$\kappa$ can be transformed away by
adding a term $-{\boldmath{E}.\nabla u}$ on the r.h.s of Eq.~\ref{eq1}.
This relates
the curvature-induced motion of a weakly curved scroll wave
\cite{keen3d,bik3d} 
to
2D spiral drift in an external field $\boldmath{E}$ and gives
for the small $k_z$ behavior of the translation
branches, 
\begin{equation}
\sigma_{\pm}(k_z)= \pm i \omega_1 +(
-\alpha_{\parallel} \pm i 
\alpha_{\perp})
k_z^2
\label{drifttrans}
\end{equation}
This relation 
between the translation modes small $k_z$ curvature and the independently
measured drift coefficients $\alpha_{\parallel}$ and $\alpha_{\perp}$
\cite{noteE} 
is very well satisfied by our numerics 
(see the captions of Fig.~\ref{bothstab},\ref{transdes},\ref{meandes}).
In the untwisted case, whether translation modes
are stabilized or destabilized by the introduction of a third dimension
is thus 
precisely related to a property of 2D spirals namely, their known change
of drift direction as kinetics parameters vary \cite{krins}.

The analogous question for meander
modes is more complex than anticipated. A
simple normal form proposed in \cite{armit} extends
Eq.(\ref{drifttrans}) to the low
$k_z$ behavior of the two complex conjugate meander branches
$\sigma_{m,\pm}(k_z)= \sigma_{m,\pm}(0)+(\alpha_{\parallel}\pm i\alpha_{\perp})
k_z^2 $.
In
contrast to Eq.~(\ref{drifttrans}), this relation  
is not obeyed by the data of Fig.~\ref{meandes}. It
even qualitatively disagrees with the data of Fig.~\ref{bothstab}a since it 
predicts that the low $k_z$ curvature of the translation and meander branches 
real parts are
of opposite signs contrary to what we find. Thus, a simple two-dimensional
account of the long-wavelength dependence of the meander
branch remains to be developed.

 For twisted scroll waves, we have found
that the mechanism of  the 'sproing' instability is a twist-induced deformation
of the translation branches.
This coupling of twist to the translation modes is
not present in analytical calculations using averaging techniques
\cite{keen3d,bik3d}. We will show elsewhere \cite{hhk} that it
can be analytically captured in the
weakly excitable limit using the techniques of\cite{hk}. Nevertheless, the 
shape of the resulting translation branches and specially their double-peak 
structure shown on Fig.4a remains to be better understood.
More generally, the precise knowledge of the important 
linear modes and their behavior,
should allow the development of  controlled reduced descriptions of  
scroll wave  nonlinear dynamics.
We
also hope that our results and others will encourage  
further experimental studies of
scroll wave instabilities.

{\em Acknowledgments} We are grateful to A. Karma for very stimulating
discussions on scroll waves. 
We also wish to thank M. E. Brachet,  H. Chat\'e, C. Nore and G. Rousseau 
for instructive discussions. Numerical computations were performed in part
at IDRIS.


\begin{thebibliography}{999}
\bibitem{noteE} In a
a small field $\boldmath{E}$ 
the spiral center of rotation drifts at a velocity $\alpha_{\parallel}
E$ along the direction of the field and $\alpha_{\perp} E$
in the perpendicular direction $\boldmath{\omega_1\times E}$ where
$\boldmath{\omega_1}$
is the spiral rotation vector. This unambiguously fixes the sign of
$\alpha_{\perp} E$ which depends on the spiral sense of rotation.
\end{thebibliography}
\end{document}